\documentclass [twocolumn,prl,a4paper,altaffilletter,superscriptaddress,amsmath,amssymb]{revtex4}
\usepackage{times}
\usepackage{graphicx}
\usepackage{bm}
\begin {document}
\title{Effect of varying material anisotropy on critical
current anisotropy in vicinal YBa$_2$Cu$_3$O$_{7-\delta}$ thin
films}
\author {J. H. Durrell}
\pacs{test} \email{jhd25@cam.ac.uk} \affiliation {Department of
Materials Science and Metallurgy and IRC in Superconductivity,
University of Cambridge, Pembroke Street, Cambridge, CB2 3QZ, UK.}
\author {R. R\"{o}ssler}
\author {M. P. Delamare}
\altaffiliation{Now at: CNRT Mat\'{e}riaux de Basse-Normandie /
ENSICAEN, 6 Boulevard Mar\'{e}chal Juin, 14050 CAEN cedex 4,
France }
\author {J. D. Pedarnig}
\author {D. B\"{a}uerle}
\affiliation {Angewandte Physik, Johannes-Kepler-Universit\"{a}t
Linz, A-4040 Linz, Austria}
\author {J. E. Evetts}
\affiliation {Department of Materials Science and Metallurgy and
IRC in Superconductivity, University of Cambridge, Pembroke
Street, Cambridge, CB2 3QZ, UK.}

\begin {abstract} The high $T_{c}$ cuprate
superconductors are noted for their anisotropic layered structure,
certain of these materials indeed tend toward the limit of a
Lawrence-Doniach superconductor. However,
YBa$_2$Cu$_3$O$_{7-\delta}$ has a smaller anisotropy than would be
expected from its interlayer spacing. This is due to the cuprate
chains in the structure. To investigate the influence of the chain
oxygen on transport properties critical current versus applied
field angle measurements were performed on fully oxygenated and
de-oxygenated YBa$_2$Cu$_3$O$_{7-\delta}$ thin films and optimally
oxygenated Y$_{0.75}$Ca$_{0.2}$Ba$_2$Cu$_3$O$_{7-\delta}$ thin
films. The films were grown on 10$^{\circ}$ mis-cut SrTiO$_3$
substrates to enable the intrinsic vortex channelling effect to be
observed. The form of the vortex channelling minimum observed in
field angle dependent critical current studies on the films was
seen to depend on film oxygenation. The vortex channelling effect
is dependent on a angular dependent cross-over to a string-pancake
flux line lattice. The results obtained appear to be consistent
with the prediction of Blatter et al. [Rev. Mod. Phys., 66 (4):
1125 (1994)] that increased superconducting anisotropy leads to
the kinked string-pancake lattice existing over a smaller angular
range.
\end {abstract}
\maketitle

Among the cuprate high $T_c$ superconductors
YBa$_2$Cu$_3$O$_{7-\delta}$ (YBCO) is of particular interest due
to its relatively low anisotropy parameter. For example while,
$d$, the cuprate plane spacing is 1.2nm in YBCO as compared to
1.5nm in BSCCO \cite{datt92}, values for the superconducting mass
anisotropy parameter $\gamma^2=(m_c/m_a)$ are quoted as 5-7 for
YBCO and 50-200 for Bi$_2$Sr$_2$Ca$_1$Cu$_2$O$_{8+x}$
\cite{far89,Chi92,Mart92,Bla94} (the large uncertainty in the
Bi$_2$Sr$_2$Ca$_1$Cu$_2$O$_{8+x}$ value is due to the difficulty
in measuring its $c$ axis transport properties).

The structure of YBCO contains 'cuprate chains' inside the
blocking layers between the cuprate planes associated with
superconductivity. In most cuprate superconductors the blocking
layers act as an insulating charge reservoir, in YBCO however, the
cuprate chains also have superconducting properties \cite{tall95}.
This is the feature of YBCO which leads to the lower anisotropy
seen in YBCO as compared to that predicted on the basis of the
interlayer spacing. The chain oxygen in YBCO is highly mobile,
under doping or over doping of the material by moving the
oxygenation away from $\delta \sim 0.08$ is associated with a
reduction in $T_c$. A reduced oxygenation level will, however,
also increase the superconducting anisotropy of the material by
disrupting the cuprate chains. Importantly the removal of oxygen
from the chains does not increase pinning \cite{Chri91,oss92}.
However, there is a definite increase in superconducting
anisotropy with $\delta$ at larger values \cite{cava90,hou94},
indeed Deak \textit{et al.} \cite{dea95} indicate that the
anisotropy more than doubles for $\delta>0.2$.

 Intrinsic vortex channelling
along the $a$-$b$ planes has previously been reported in optimally
doped YBCO films grown on mis-cut (vicinal) substrates
\cite{ber97}. The development of the channelling minima in these
measurements has been attributed to the appearance of a kinked
(string/pancake) vortex state as the angle, $\theta$, between the
$a$-$b$ planes and the applied field is reduced. The pinning of
vortex string elements against a Lorentz force directed parallel
to the $a$-$b$ planes is relatively weak. By using a vicinal
geometry thin film it is possible to arrange for a component of
the Lorentz force to act along the $a$-$b$ planes. This leads to a
marked reduction in the critical current when the magnetic field
is applied parallel to the $a$-$b$ planes.
\begin{figure}
\includegraphics{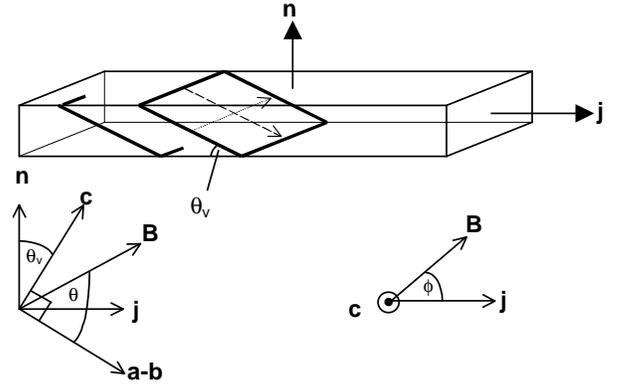}% Here is how to import EPS art
\caption{\label{fig:1} Measurement geometry. The angle $\theta$ is
that between the applied magnetic field and the $a$-$b$
crystallographic planes. The angle $\phi$ is that of the rotation
of the applied field with respect to the direction of vicinal
tilt, the magnitude of which is defined as $\theta_v$. The
measurements reported here were carried out with $\phi=0$ so the
directions of the normal to the film, \textbf{n}, the $c$-axis,
the field and the current all lay in the same plane.}
\end{figure}
The kinked vortex state is intimately related to the anisotropy of
the superconducting material. The variation of the structure of
flux lines in YBCO as the angle, $\theta$, between the applied
field and the $a$-$b$ planes varies has been discussed by Blatter
et al. \cite{Bla94}. For temperatures below $T_{cr}$, where
$T_{cr}$ is defined by $\epsilon\xi_{ab}(T_{cr})=d/\surd2$
($\sim$80K in optimally doped YBCO), the critical angle between
the applied magnetic field and the $a$-$b$ planes required for the
development of a fully kinked string-pancake structure is given
by:
\begin{equation}
\label{eq:1} \tan(\theta_2)=\epsilon
\end{equation}
where $\epsilon=1/\gamma$ is the Ginzburg-Landau (G-L) anisotropy
parameter. Between $\theta_2$ and a larger angle $\theta_1$ the
flux lines exhibit a distorted structure while for
$\theta>\theta_1$ they adopt the conventional rectilinear
Abrikosov form. $\theta_1$ is defined as
$\tan(\theta_1)=d/\epsilon_{ab}(T/T_c)$. From this it can be
deduced that an increasing G-L anisotropy parameter is predicted
to lead to a smaller angular range of kinked structure.

In order to study the effect of material anisotropy on the vortex
channelling effect vicinal YBCO films with different anisotropies
were prepared. The YBa$_2$Cu$_3$O$_{7-\delta}$ and
Y$_{0.75}$Ca$_{0.2}$Ba$_2$Cu$_3$O$_{7-\delta}$ thin films were
prepared under identical conditions on 10$^\circ$ miscut vicinal
(001) SrTiO$_3$ substrates by pulsed-laser deposition
\cite{bau00}. The film were $\sim 200nm$ thick. In situ
post-annealing of the films wa s performed at an oxygen pressure
of 800 mbar and 0.7 mbar for the fully oxygenated and
de-oxygenated YBCO films respectively. The Ca-substited film was
annealed at $\sim$ 15 mbar for optimum critical temperature
\cite{dela02}. Tracks were patterned onto the films using
photolithography and Ar ion milling to give tracks 100$\mu$m by
10$\mu$m both parallel, \textsf{\textbf{L}}, and perpendicular,
 \textsf{\textbf{T}}, to the vicinal steps.
Critical current values were determined from IV characteristics
using a 0.55$\mu$V criterion, measurements were performed using a
two-axis goniometer mounted in a 8T magnet. The experimental
geometry is shown in Figure \ref{fig:1}.
\begin{figure}
\includegraphics{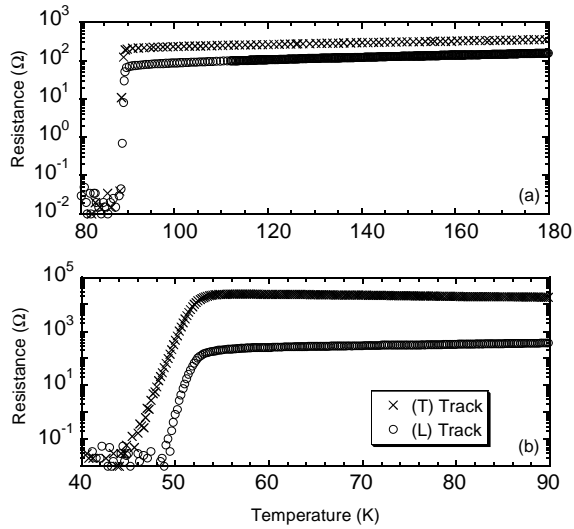}
\caption{\label{fig:2} Resistance versus temperature plots for the
fully (a) and de-oxygenated (b) YBCO films. It can be seen that
the in-plane resistivity anisotropy is greatly increased in the
de-oxygenated film.}
\end{figure}
The resistive transition for both films was measured on both the
\textsf{\textbf{T}} and \textsf{\textbf{L}} tracks. As discussed
by Zahner et al. \cite{zah98} it is possible to deduce the
resistance anisotropy between the $a$-$b$ and the $c$ axes from
the resistivity ratio between the \textsf{\textbf{T}} and
\textsf{\textbf{L}} tracks, $\gamma_{TL}$:
$\gamma_\rho=[\gamma_{TL}-\cos^2(\theta_v)]/\sin^2(\theta_v)$,
where $\theta_v$ is the vicinal mis-cut angle. The resistance
temperature data shown in Fig. \ref{fig:2} indicates a resistivity
anisotropy at 100K of 66 for the optimally oxygenated film and
1600 for the de-oxygenated film. These values are consistent with
those observed in single crystal samples \cite{datt92}.

The $T_c$ values were 90 K and 51 K for the fully- and
de-oxygenated films respectively. Although other factors apart
from $\delta$ affect $T_c$ this value of $T_c$ suggests a $\delta$
value of about 0.4 \cite{cyro95} for the de-oxygenated film. From
Eq. \ref{eq:1} and the relationship between $\delta$ and $\gamma$
found by Deak \textit{et al.} \cite{dea95} this would predict, at
$t$=0, a value for $\theta_2$ of 3$^{\circ}$  as compared to
11$^{\circ}$ for fully oxygenated YBCO. Furthermore $T_{cr}$
becomes closer to $T_c$ with increasing anisotropy, this means the
channelling minimum should be visible in $J_c(\theta)$
measurements over a wider range of reduced temperature in the
de-oxygenated sample.

\begin{figure}
\includegraphics{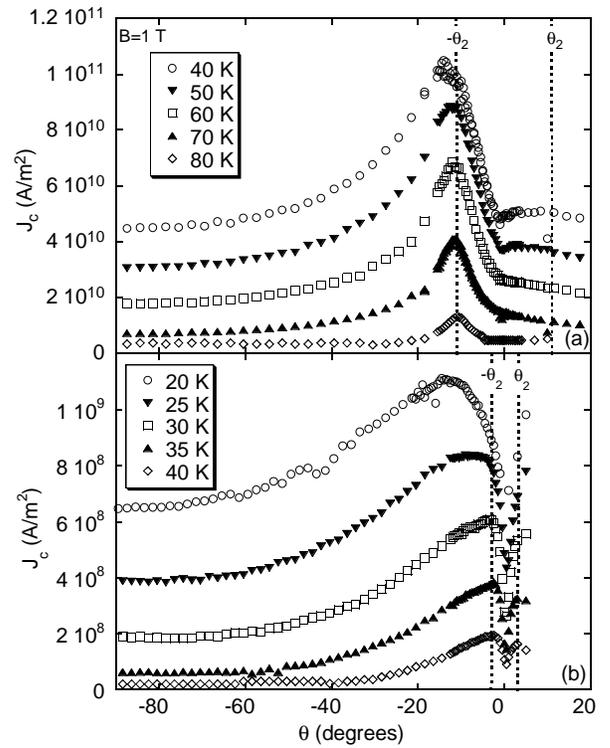}
\caption{\label{fig:3} $J_c(\theta)$ plots for the fully (a) and
de-oxygenated (b) films measured. These data were obtained in the
$\phi$=0 geometry so the channelling minima and the 'force-free'
peak are offset by the vicinal angle, 10$^\circ$. The dotted lines
indicate the predicted values of $\theta_2$.}
\end{figure}

Figure \ref{fig:3} shows the form of the vortex channelling effect
at a magnetic field of 1 Tesla and various temperatures for the
fully oxygenated and de-oxygenated films measured. The results
obtained for the fully oxygenated sample are, as would be
expected, very similar to those previously reported \cite{ber97}.
From the data presented in Fig. \ref{fig:3} it is clear that there
is a distinct change in the form of the channelling minima for the
de-oxygenated film. As would be expected the angular range over
which a departure from the conventional truncated $1/cos \theta$
force free behavior appears to be smaller for the case of the
de-oxygenated film. As the films were grown under identical
conditions, apart from the oxygen partial pressure during
annealing,  this change in the $J_c(\theta)$ behavior is
reasonably associated with a change in the anisotropy of the
sample studied.

Calcium doping has been extensively studied as a means of
improving the properties of grain boundaries in YBCO. In the
grain, however, in order to obtain a high $T_c$ a reduced chain
oxygen occupancy is required as the Ca doping shifts the peak in
the $T_c$ versus $\delta$ curve. 25\% Ca-substituted YBCO with an
oxygenation corresponding to optimum $T_c$ will therefore have
oxygen vacancies in its cuprate chains and would be expected
therefore to exhibit a stronger superconducting anisotropy.

\begin{figure}
\includegraphics{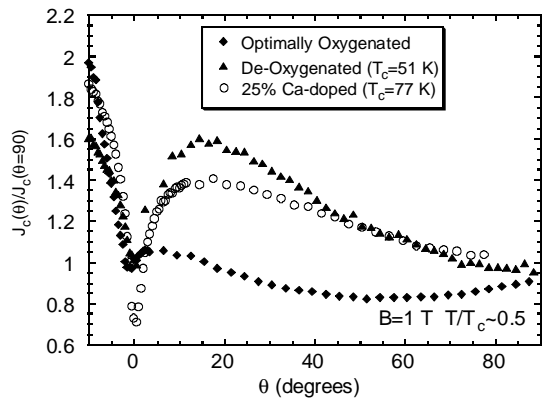}
\caption{\label{fig:4} $J_c(\theta)/J_c(\theta=90)$ data plotted
for all three films. These data was taken at 1 T and the same
reduced temperature.}
\end{figure}

A comparison of the $J_c(\theta)$ behavior of a Ca-doped film with
a fully- and de-oxygenated film is shown in Figure \ref{fig:4}.
The Ca-doped film was grown in the same PLD system and under
similar conditions, but with a 25\% Ca-substituted target. It is
interesting to note that the Ca-doped film exhibits broadly
similar behavior to the de-oxygenated film. This is evidence that
Ca-doped YBCO also exhibits an increased superconducting
anisotropy and that this is due to the oxygen vacancies in the
cuprate chain needed to obtain an optimum $T_c$.
From the measurements presented in this letter it is clear that
the oxygenation of the cuprate planes strongly affects the
superconducting anisotropy of the material. A variation of this
anisotropy changes the way the structure of individual flux lines
evolves as the applied magnetic field is rotated with respect to
the cuprate planes. This anisotropy dependent change is reflected
in the vicinal channelling effect observed in YBCO samples with
differing chain oxygen concentration. This increased anisotropy is
also apparent in Ca-doped YBCO which, for optimum $T_c$, must have
a lower oxygenation than non Ca-doped YBCO.
\bibliography {de-ox}

\begin{thebibliography}{16}
\expandafter\ifx\csname natexlab\endcsname\relax\def\natexlab#1{#1}\fi
\expandafter\ifx\csname bibnamefont\endcsname\relax
  \def\bibnamefont#1{#1}\fi
\expandafter\ifx\csname bibfnamefont\endcsname\relax
  \def\bibfnamefont#1{#1}\fi
\expandafter\ifx\csname citenamefont\endcsname\relax
  \def\citenamefont#1{#1}\fi
\expandafter\ifx\csname url\endcsname\relax
  \def\url#1{\texttt{#1}}\fi
\expandafter\ifx\csname urlprefix\endcsname\relax\def\urlprefix{URL }\fi
\providecommand{\bibinfo}[2]{#2}
\providecommand{\eprint}[2][]{\url{#2}}

\bibitem[{\citenamefont{Datta}(1992)}]{datt92}
\bibinfo{author}{\bibfnamefont{T.}~\bibnamefont{Datta}}, in
  \emph{\bibinfo{booktitle}{Concise Encylopedia of Magnetic and Superconducting
  Materials}}, edited by \bibinfo{editor}{\bibfnamefont{J.~E.}
  \bibnamefont{Evetts}} (\bibinfo{publisher}{Pergamon},
  \bibinfo{address}{Oxford}, \bibinfo{year}{1992}).

\bibitem[{\citenamefont{Farrell et~al.}(1989)\citenamefont{Farrell, Bonham,
  Foster, Chang, Jiang, Vandervoort, Lam, and Kogan}}]{far89}
\bibinfo{author}{\bibfnamefont{D.~E.} \bibnamefont{Farrell}},
  \bibinfo{author}{\bibfnamefont{S.}~\bibnamefont{Bonham}},
  \bibinfo{author}{\bibfnamefont{J.}~\bibnamefont{Foster}},
  \bibinfo{author}{\bibfnamefont{Y.~C.} \bibnamefont{Chang}},
  \bibinfo{author}{\bibfnamefont{P.~Z.} \bibnamefont{Jiang}},
  \bibinfo{author}{\bibfnamefont{K.~G.} \bibnamefont{Vandervoort}},
  \bibinfo{author}{\bibfnamefont{D.~J.} \bibnamefont{Lam}}, \bibnamefont{and}
  \bibinfo{author}{\bibfnamefont{V.~G.} \bibnamefont{Kogan}},
  \bibinfo{journal}{Physical Review Letters} \textbf{\bibinfo{volume}{63}},
  \bibinfo{pages}{782} (\bibinfo{year}{1989}).

\bibitem[{\citenamefont{Chikumoto et~al.}(1992)\citenamefont{Chikumoto,
  Konczykowski, Motohira, and Kishio}}]{Chi92}
\bibinfo{author}{\bibfnamefont{N.}~\bibnamefont{Chikumoto}},
  \bibinfo{author}{\bibfnamefont{M.}~\bibnamefont{Konczykowski}},
  \bibinfo{author}{\bibfnamefont{N.}~\bibnamefont{Motohira}}, \bibnamefont{and}
  \bibinfo{author}{\bibfnamefont{K.}~\bibnamefont{Kishio}},
  \bibinfo{journal}{Physica C} \textbf{\bibinfo{volume}{199}},
  \bibinfo{pages}{32} (\bibinfo{year}{1992}).

\bibitem[{\citenamefont{Martinez et~al.}(1992)\citenamefont{Martinez,
  Brongersma, Koshelev, Ivlev, Kes, Griessen, de~Groot, Tarnavski, and
  Menovsky}}]{Mart92}
\bibinfo{author}{\bibfnamefont{J.~C.} \bibnamefont{Martinez}},
  \bibinfo{author}{\bibfnamefont{S.~H.} \bibnamefont{Brongersma}},
  \bibinfo{author}{\bibfnamefont{A.}~\bibnamefont{Koshelev}},
  \bibinfo{author}{\bibfnamefont{B.}~\bibnamefont{Ivlev}},
  \bibinfo{author}{\bibfnamefont{P.~H.} \bibnamefont{Kes}},
  \bibinfo{author}{\bibfnamefont{R.~P.} \bibnamefont{Griessen}},
  \bibinfo{author}{\bibfnamefont{D.~G.} \bibnamefont{de~Groot}},
  \bibinfo{author}{\bibfnamefont{Z.}~\bibnamefont{Tarnavski}},
  \bibnamefont{and} \bibinfo{author}{\bibfnamefont{A.~A.}
  \bibnamefont{Menovsky}}, \bibinfo{journal}{Physical Review Letters}
  \textbf{\bibinfo{volume}{69}}, \bibinfo{pages}{2276} (\bibinfo{year}{1992}).

\bibitem[{\citenamefont{Blatter et~al.}(1994)\citenamefont{Blatter, Feigelman,
  Geshkenbein, Larkin, and Vinokur}}]{Bla94}
\bibinfo{author}{\bibfnamefont{G.}~\bibnamefont{Blatter}},
  \bibinfo{author}{\bibfnamefont{M.~V.} \bibnamefont{Feigelman}},
  \bibinfo{author}{\bibfnamefont{V.~B.} \bibnamefont{Geshkenbein}},
  \bibinfo{author}{\bibfnamefont{A.~I.} \bibnamefont{Larkin}},
  \bibnamefont{and} \bibinfo{author}{\bibfnamefont{V.~M.}
  \bibnamefont{Vinokur}}, \bibinfo{journal}{Reviews of Modern Physics}
  \textbf{\bibinfo{volume}{66}}, \bibinfo{pages}{1125} (\bibinfo{year}{1994}).

\bibitem[{\citenamefont{Tallon et~al.}(1995)\citenamefont{Tallon, Bernhard,
  Binninger, Hofer, Williams, Ansaldo, Budnick, and Niedermayer}}]{tall95}
\bibinfo{author}{\bibfnamefont{J.~L.} \bibnamefont{Tallon}},
  \bibinfo{author}{\bibfnamefont{C.}~\bibnamefont{Bernhard}},
  \bibinfo{author}{\bibfnamefont{U.}~\bibnamefont{Binninger}},
  \bibinfo{author}{\bibfnamefont{A.}~\bibnamefont{Hofer}},
  \bibinfo{author}{\bibfnamefont{G.~V.~M.} \bibnamefont{Williams}},
  \bibinfo{author}{\bibfnamefont{E.~J.} \bibnamefont{Ansaldo}},
  \bibinfo{author}{\bibfnamefont{J.~I.} \bibnamefont{Budnick}},
  \bibnamefont{and}
  \bibinfo{author}{\bibfnamefont{C.}~\bibnamefont{Niedermayer}},
  \bibinfo{journal}{Physical Review Letters} \textbf{\bibinfo{volume}{74}},
  \bibinfo{pages}{1008} (\bibinfo{year}{1995}).

\bibitem[{\citenamefont{Christen and Feenstra}(1991)}]{Chri91}
\bibinfo{author}{\bibfnamefont{D.~K.} \bibnamefont{Christen}} \bibnamefont{and}
  \bibinfo{author}{\bibfnamefont{R.}~\bibnamefont{Feenstra}},
  \bibinfo{journal}{Physica C} \textbf{\bibinfo{volume}{185}},
  \bibinfo{pages}{2225} (\bibinfo{year}{1991}).

\bibitem[{\citenamefont{Ossandon et~al.}(1992)\citenamefont{Ossandon, Thompson,
  Christen, Sales, Kerchner, Thomson, Sun, Lay, and Tkaczyk}}]{oss92}
\bibinfo{author}{\bibfnamefont{J.~G.} \bibnamefont{Ossandon}},
  \bibinfo{author}{\bibfnamefont{J.~R.} \bibnamefont{Thompson}},
  \bibinfo{author}{\bibfnamefont{D.~K.} \bibnamefont{Christen}},
  \bibinfo{author}{\bibfnamefont{B.~C.} \bibnamefont{Sales}},
  \bibinfo{author}{\bibfnamefont{H.~R.} \bibnamefont{Kerchner}},
  \bibinfo{author}{\bibfnamefont{J.~O.} \bibnamefont{Thomson}},
  \bibinfo{author}{\bibfnamefont{Y.~R.} \bibnamefont{Sun}},
  \bibinfo{author}{\bibfnamefont{K.~W.} \bibnamefont{Lay}}, \bibnamefont{and}
  \bibinfo{author}{\bibfnamefont{J.~E.} \bibnamefont{Tkaczyk}},
  \bibinfo{journal}{Physical Review B} \textbf{\bibinfo{volume}{45}},
  \bibinfo{pages}{12534} (\bibinfo{year}{1992}).

\bibitem[{\citenamefont{Cava et~al.}(1990)\citenamefont{Cava, Hewat, Hewat,
  Batlogg, Marezio, Rabe, Krajewski, Peck, and Rupp}}]{cava90}
\bibinfo{author}{\bibfnamefont{R.~J.} \bibnamefont{Cava}},
  \bibinfo{author}{\bibfnamefont{A.~W.} \bibnamefont{Hewat}},
  \bibinfo{author}{\bibfnamefont{E.~A.} \bibnamefont{Hewat}},
  \bibinfo{author}{\bibfnamefont{B.}~\bibnamefont{Batlogg}},
  \bibinfo{author}{\bibfnamefont{M.}~\bibnamefont{Marezio}},
  \bibinfo{author}{\bibfnamefont{K.~M.} \bibnamefont{Rabe}},
  \bibinfo{author}{\bibfnamefont{J.~J.} \bibnamefont{Krajewski}},
  \bibinfo{author}{\bibfnamefont{W.~F.} \bibnamefont{Peck}}, \bibnamefont{and}
  \bibinfo{author}{\bibfnamefont{L.~W.} \bibnamefont{Rupp}},
  \bibinfo{journal}{Physica C} \textbf{\bibinfo{volume}{165}},
  \bibinfo{pages}{419} (\bibinfo{year}{1990}).

\bibitem[{\citenamefont{Hou et~al.}(1994)\citenamefont{Hou, Deak, Metcalf, and
  McElfresh}}]{hou94}
\bibinfo{author}{\bibfnamefont{L.~F.} \bibnamefont{Hou}},
  \bibinfo{author}{\bibfnamefont{J.}~\bibnamefont{Deak}},
  \bibinfo{author}{\bibfnamefont{P.}~\bibnamefont{Metcalf}}, \bibnamefont{and}
  \bibinfo{author}{\bibfnamefont{M.}~\bibnamefont{McElfresh}},
  \bibinfo{journal}{Physical Review B} \textbf{\bibinfo{volume}{50}},
  \bibinfo{pages}{7226} (\bibinfo{year}{1994}).

\bibitem[{\citenamefont{Deak et~al.}(1995)\citenamefont{Deak, Hou, Metcalf, and
  McElfresh}}]{dea95}
\bibinfo{author}{\bibfnamefont{J.}~\bibnamefont{Deak}},
  \bibinfo{author}{\bibfnamefont{L.~F.} \bibnamefont{Hou}},
  \bibinfo{author}{\bibfnamefont{P.}~\bibnamefont{Metcalf}}, \bibnamefont{and}
  \bibinfo{author}{\bibfnamefont{M.}~\bibnamefont{McElfresh}},
  \bibinfo{journal}{Physical Review B} \textbf{\bibinfo{volume}{51}},
  \bibinfo{pages}{705} (\bibinfo{year}{1995}).

\bibitem[{\citenamefont{Berghuis et~al.}(1997)\citenamefont{Berghuis,
  DiBartolomeo, Wagner, and Evetts}}]{ber97}
\bibinfo{author}{\bibfnamefont{P.}~\bibnamefont{Berghuis}},
  \bibinfo{author}{\bibfnamefont{E.}~\bibnamefont{DiBartolomeo}},
  \bibinfo{author}{\bibfnamefont{G.~A.} \bibnamefont{Wagner}},
  \bibnamefont{and} \bibinfo{author}{\bibfnamefont{J.~E.}
  \bibnamefont{Evetts}}, \bibinfo{journal}{Physical Review Letters}
  \textbf{\bibinfo{volume}{79}}, \bibinfo{pages}{2332} (\bibinfo{year}{1997}).

\bibitem[{\citenamefont{B\"{a}uerle}(2000)}]{bau00}
\bibinfo{author}{\bibfnamefont{D.}~\bibnamefont{B\"{a}uerle}},
  \emph{\bibinfo{title}{Laser Processing and Chemistry}}
  (\bibinfo{publisher}{Springer}, \bibinfo{address}{New York},
  \bibinfo{year}{2000}), \bibinfo{edition}{3rd} ed.

\bibitem[{\citenamefont{Delamare et~al.}(2002)\citenamefont{Delamare,
  Sch\"{o}ppl, Pedarnig, and B\"{a}uerle}}]{dela02}
\bibinfo{author}{\bibfnamefont{M.~P.} \bibnamefont{Delamare}},
  \bibinfo{author}{\bibfnamefont{R.~K.} \bibnamefont{Sch\"{o}ppl}},
  \bibinfo{author}{\bibfnamefont{J.}~\bibnamefont{Pedarnig}}, \bibnamefont{and}
  \bibinfo{author}{\bibfnamefont{D.}~\bibnamefont{B\"{a}uerle}},
  \bibinfo{journal}{Physica C} \textbf{\bibinfo{volume}{372-376}},
  \bibinfo{pages}{638} (\bibinfo{year}{2002}).

\bibitem[{\citenamefont{Zahner et~al.}(1998)\citenamefont{Zahner, Stierstorfer,
  R\"{o}ssler, Pedarnig, B\"{a}uerle, and Lengfellner}}]{zah98}
\bibinfo{author}{\bibfnamefont{T.}~\bibnamefont{Zahner}},
  \bibinfo{author}{\bibfnamefont{R.}~\bibnamefont{Stierstorfer}},
  \bibinfo{author}{\bibfnamefont{R.}~\bibnamefont{R\"{o}ssler}},
  \bibinfo{author}{\bibfnamefont{J.~D.} \bibnamefont{Pedarnig}},
  \bibinfo{author}{\bibfnamefont{D.}~\bibnamefont{B\"{a}uerle}},
  \bibnamefont{and}
  \bibinfo{author}{\bibfnamefont{H.}~\bibnamefont{Lengfellner}},
  \bibinfo{journal}{Physica C} \textbf{\bibinfo{volume}{298}},
  \bibinfo{pages}{91} (\bibinfo{year}{1998}).

\bibitem[{\citenamefont{Cyrot and Pavuna}(1995)}]{cyro95}
\bibinfo{author}{\bibfnamefont{M.}~\bibnamefont{Cyrot}} \bibnamefont{and}
  \bibinfo{author}{\bibfnamefont{D.}~\bibnamefont{Pavuna}},
  \emph{\bibinfo{title}{Introduction to Superconductivity and High-Tc
  materials}} (\bibinfo{publisher}{World Scientific},
  \bibinfo{address}{Singapore}, \bibinfo{year}{1995}).

\end{thebibliography}
\end{document}